\documentclass[conference]{IEEEtran}
\IEEEoverridecommandlockouts
\usepackage{cite}
\usepackage{amsmath,amssymb,amsfonts}
\usepackage{algorithmic}
\usepackage{graphicx}
\usepackage{textcomp}
\usepackage{xcolor}
\usepackage{hyperref}
\def\BibTeX{{\rm B\kern-.05em{\sc i\kern-.025em b}\kern-.08em
    T\kern-.1667em\lower.7ex\hbox{E}\kern-.125emX}}

\title{IoT and Man-in-the-Middle Attacks}

\author{\IEEEauthorblockN{Hamidreza Fereidouni}
\IEEEauthorblockA{\textit{Department of Computer Science}\\
\textit{and Operations Research}\\
\textit{University of Montreal}\\
Montreal, QC, Canada\\
hamidreza.fereidouni@umontreal.ca}
\and
\IEEEauthorblockN{Olga Fadeitcheva}
\IEEEauthorblockA{\textit{Department of Computer Engineering}\\
\textit{and Software Engineering}\\
\textit{Polytechnique Montreal}\\
Montreal, QC, Canada\\
olga.fadeitcheva@polymtl.ca}
\and
\IEEEauthorblockN{Mehdi Zalai}
\IEEEauthorblockA{\textit{Department of Computer Engineering}\\
\textit{and Software Engineering}\\
\textit{Polytechnique Montreal}\\
Montreal, QC, Canada\\
mehdi.zalai@polymtl.ca}
}

\begin{document}

\maketitle

\begin{abstract}
This paper provides an overview of the Internet of Things (IoT) and its significance. It discusses the concept of Man-in-the-Middle (MitM) attacks in detail, including their causes, potential solutions, and challenges in detecting and preventing such attacks. The paper also addresses the current issues related to IoT security and explores future methods and facilities for improving detection and prevention mechanisms against MitM.
\end{abstract}

\begin{IEEEkeywords}
IoT, Internet of Things, IoT Security, MitM, Man-in-the-Middle, MitM Detection, MitM Prevention
\end{IEEEkeywords}

\section{Introduction}
\label{sec:Introduction}

Today, the internet plays a pivotal role in our lives. The internet can be primitively considered a network of networks; also, it is technically defined as a unified, interconnected system of computer networks, from educational computer networks to governmental ones, which operate on predefined, accepted networking protocols \cite{b1}. Conventionally, we use the internet to do our daily routines, from official correspondence and private communications to buying goods and banking operations. The Internet of Things (IoT) is defined as a network of physical objects comprising sensors and actuators, and connections, which allows these things to connect and trade data through the internet \cite{b2}. This concept, independent of this name, has been around for more than two decades since the unveiling of the internet publicly.

IoT has facilitated our modern life, from industries to our personal life. For instance, a smart home utilizes IoT to provide residents with a better experience of living. The fundamental goal of IoT innovations is to provide connectivity and improve the quality of life (QoL) for clients. On a hot summer day, a smart home can determine if its resident is returning home, check the temperature and lighting level, and then make the home ready for the resident by turning on the air conditioners and the necessary lamps. Therefore, it can simply help us in energy saving and having a suitable condition simultaneously. Given this example, we readily find the term IoT as an augmented utility of the internet, which is growing rapidly in numbers and applications. Regarding what has hitherto been described, the world has been witness to a burgeoning ecosystem of IoT in terms of economics and the number of connected devices. Based on the IoT Analytics reports, the global market size is predicted to grow 19\% annually and reach around \$483 billion in spending on enterprise IoT technologies by 2027 \cite{b3}. Besides, IoT Analytics mentions there will be 34.2 billion devices connected to the internet, of which 21.5 billion will be IoT-based devices by 2025 \cite{b4}. This might be a conservative prediction, whereas there are many other predictions that show a significantly higher number of connected devices.

Internet users are witnessing a remarkable increase in the types and the number of security and privacy issues when facing these emerging, fast-growing technologies. Due to the number of connected devices and volume of data exchanged among these devices, data leakage, authentication, and identification could be named as some of the major security vulnerabilities in the Internet of Things. This matter becomes substantial when security analysts realize that approximately 98\% of all traffic among IoT devices is unencrypted \cite{b5}. Therefore, there are well-known threats and attacks, such as Distributed Denial of Service (DDoS) and Man-in-the-Middle (MitM), which are commonly able to compromise IoT-based systems. DDoS is the most popular attack because of ease of launch and no need for complex, time-consuming exploitations \cite{b6}. DDoS is a security threat that can mainly disturb the availability of services. MitM can overshadow all security and privacy facets of a system. MitM attacks are usually more complex than other attacks and difficult to identify \cite{b7}. MitM generally includes a wide variety of attacks in which a perpetrator is positioned in the middle of a communication to take control of the communication channel and reorients the original connection link (as shown in Figure 1) so as to intercept and/or alter trading data between them \cite{b8}.

\begin{figure}[htbp]
\centerline{\includegraphics[width=0.49\textwidth]{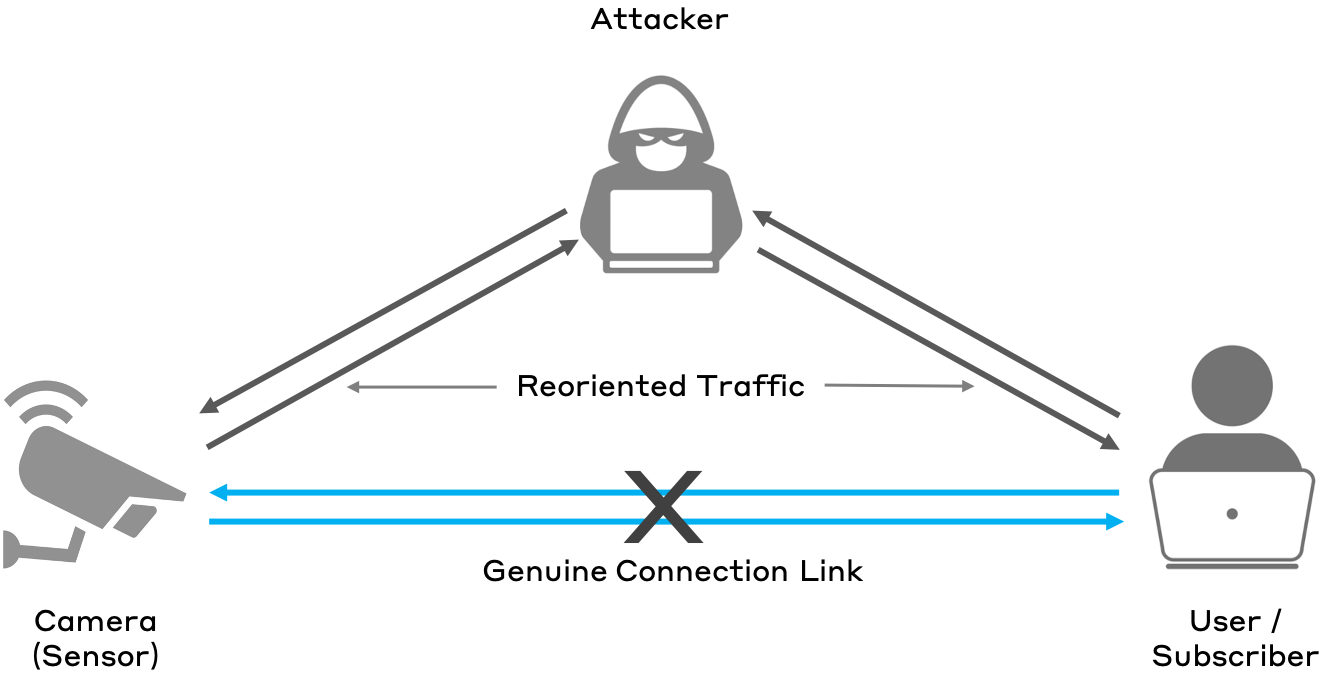}}
\caption{A General Schema of Man-in-the-Middle Attack}
\label{fig}
\end{figure}

IoT and its innovative applications penetrate every facet of human life, from industrial automation to novel medical technologies. Although IoT is a widely discussed and disruptive concept in the current internet landscape, there is currently no universally accepted security standard or framework among most businesses, despite conventional security standards that could be adapted for IoT-based environments \cite{b9}. However, there are some proposed best practices such as OWASP \cite{b36, b37}, IoT Security Maturity Model (by Industry IoT Consortium) \cite{b39}, and NIST Cybersecurity for IoT Program \cite{b38}. This situation poses a significant challenge for ensuring the security and privacy of IoT systems due to the heterogeneity of devices and protocols, as well as the wide variety of use cases that IoT-based systems can support. In these environments, there are various types of machines, sensors, and actuators with their authentic situation, protocol, and operation. These define a network with so many different nodes trading numerous transactions. This heterogeneous situation makes business bodies vulnerable to many various threats.

Under NetCraft’s report, 95\% of secured HTTP servers are vulnerable even to primitive Man-in-the-Middle attacks \cite{b10}. These kinds of reports increase the significance of MitM attacks, mainly when they occur in IoT environments where a large amount of sensitive data may be exchanged among unsecured, heterogeneous devices. Also, in many cases, MitM is intrinsically considered an Advanced Persistent Threat (APT) because of the difficulty of detection. According to this, many regular and commonly-used Intrusion Detection Systems (IDS) even cannot differentiate and identify MitM properly \cite{b11}. Thus, Man-in-the-Middle can be practically named one of the most dangerous threats in telecommunication and computer networks, which influence both the security and privacy sides of a system.

This paper explores proactive security measures and methods of identifying Man-in-the-Middle (MitM) attacks in IoT environments. It highlights important considerations to maximize the prevention of MitM in IoT systems. First, \hyperref[sec:Background and Theory]{in the second section}, the paper provides a general overview of IoT and MitM attacks. Second, \hyperref[sec:A Comparative analysis of the current state of IoT and MitM attacks]{in the fourth section}, it delves into the vulnerabilities in IoT environments, analyzes MitM attacks in these settings, and examines current prevention techniques. Finally, \hyperref[sec:Open issues]{in the fifth section}, the paper discusses future trends and challenges in the field.
 
\section{Background and Theory}
\label{sec:Background and Theory}

\subsection{Description of IoT Devices and Architecture}

The Internet of Things is characterized by connected devices that can collect and transmit data. The Subscriber-Publisher model, implemented in the Message Queuing Telemetry Transport (MQTT) protocol \cite{b12}, is a common approach in IoT. Publishers, such as sensors, generate and send data to other devices, while subscribers receive data from publishers and can be devices like actuators or users. This model enables efficient data exchange and communication between devices in IoT applications, resulting in the main four types of devices in IoT:

\begin{itemize}
    \item {Sensors:} IoT devices rely heavily on sensors to collect and transmit data from the physical world, enabling monitoring, control, automation, services, and application. IoT sensors mostly transmit data wirelessly using protocols like Wi-Fi, Bluetooth, and Zigbee.
    \item {Actuators:} IoT actuators convert digital commands into physical actions, enabling IoT devices to interact with the physical world.
    \item {Broker:} An IoT broker or gateway serves as a mediator between IoT devices and cloud-based services, performing tasks like data filtering, translation, security, and storage. Brokers enable IoT devices to communicate with cloud-based platforms and facilitate local processing and analysis of IoT data.
    \item {Users’ devices:} Users are individuals who interact with an IoT system to access and control its services or applications through different interfaces. They play a crucial role in the successful adoption and operation of IoT systems by providing feedback and insights.
\end{itemize}

\begin{figure}[htbp]
\centerline{\includegraphics[width=0.49\textwidth]{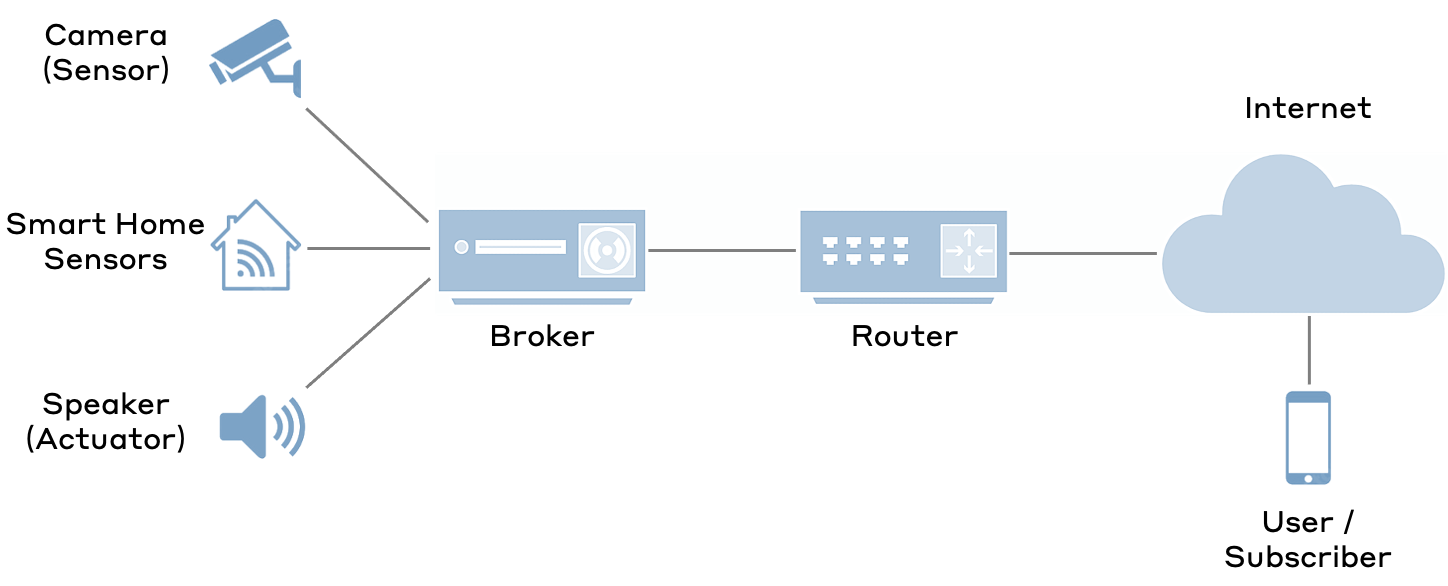}}
\caption{A General View of IoT Devices}
\label{fig}
\end{figure}

Figure 2 illustrates the devices commonly found in an IoT system, such as sensors (e.g., security cameras, smart home sensors), actuators (e.g., the speaker), and brokers that connect the elements. The figure also shows a router for internet connectivity and a user interface for remote monitoring and control.

Another substantial subject in IoT is communication methods. IoT devices communicate with each other using networking protocols across different layers. To comprehend the IoT domain, it is essential to define the constituent layers and elements of IoT and use them to delineate the potential IoT architectures that are aligned with the requisite services and fields. The IoT architecture is often divided into 3, 4, or 5 layers \cite{b13}, but from a computer networking perspective, the 3-layer model is more relevant, consisting of the application layer, network layer, and perception layer. Based on the inter-networking TCP/IP protocol stack and OSI model, as shown in Figure 3, IoT protocols can be categorized into four logical layers \cite{b14}.

\begin{figure}[htbp]
\centerline{\includegraphics[width=0.49\textwidth]{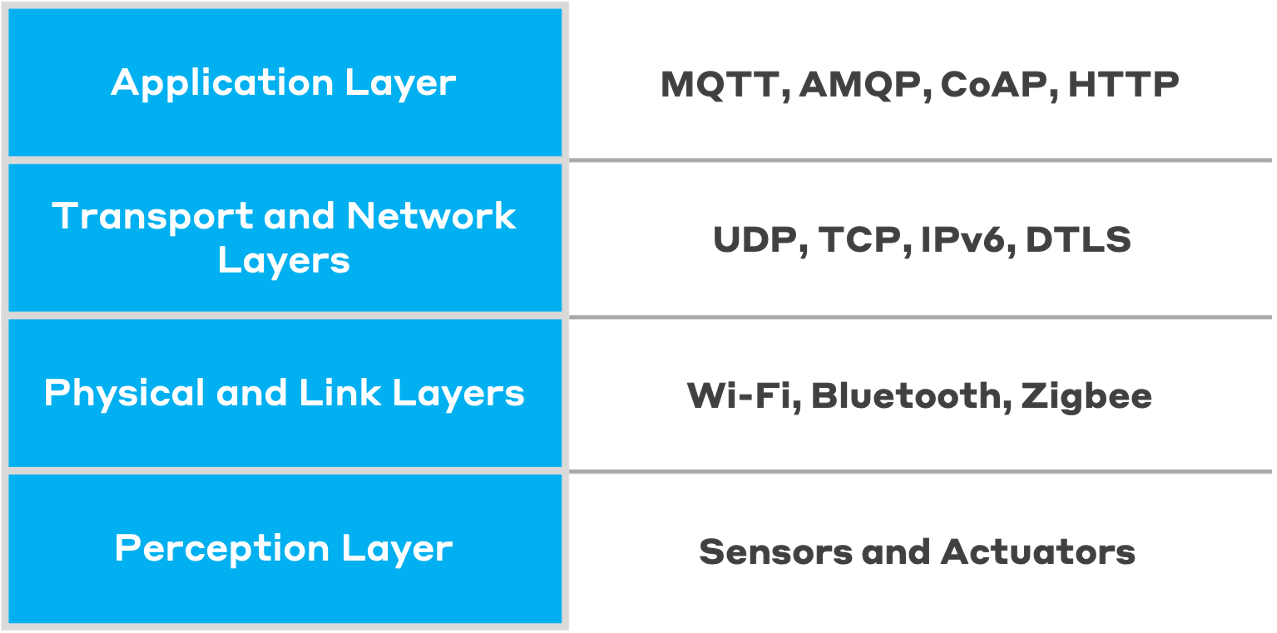}}
\caption{IoT Network Architecture Layers \cite{b14}}
\label{fig}
\end{figure}

\break

Hitherto, the types of devices and layers of protocols in IoT have been briefly described. However, one crucial aspect remains undefined. It is essential for security specialists to understand the key differences between traditional networks and IoT when dealing with IoT systems \cite{b15}. This knowledge can help in designing secure IoT systems, as there are inherent attributes and constraints in IoT that set it apart from traditional networks. Figure 4 illustrates some of the significant differences between the two.

\begin{figure}[htbp]
\centerline{\includegraphics[width=0.49\textwidth]{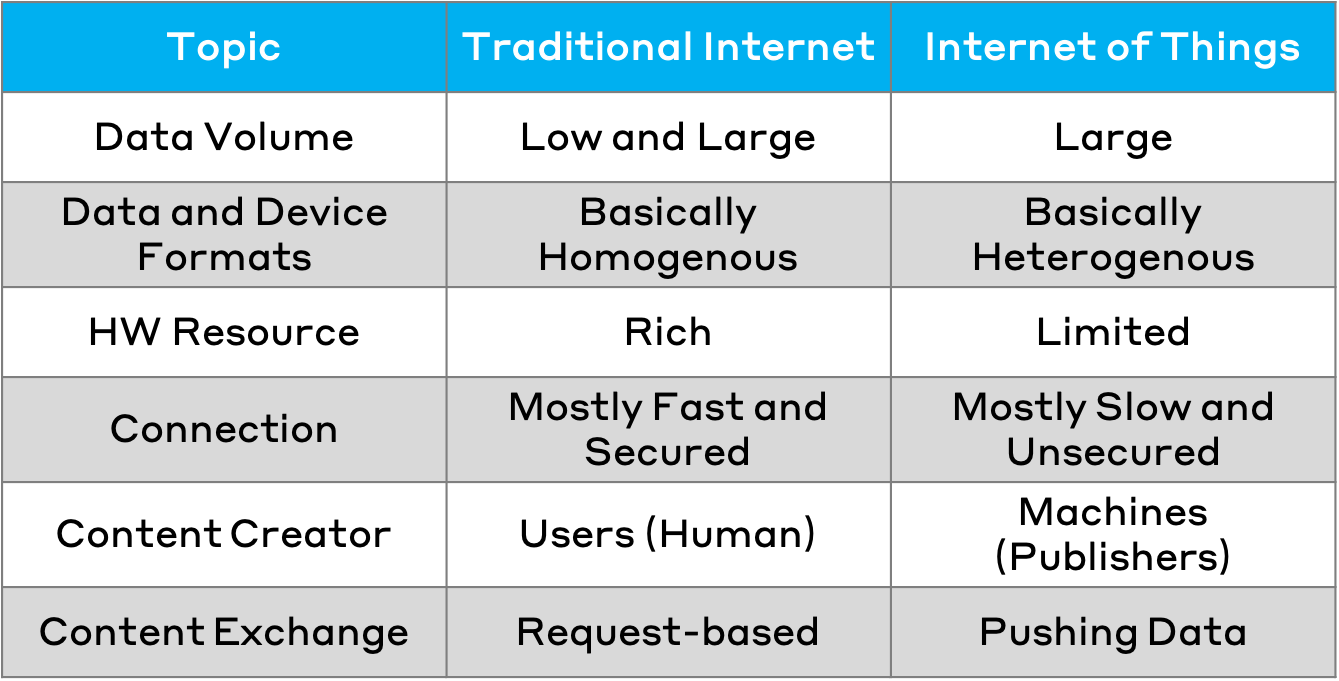}}
\caption{Traditional Internet vs. Internet of Things \cite{b15}}
\label{fig}
\end{figure}

The first three rows of the table above are particularly crucial in an IoT system. This is due to the prevalence of large data transactions, resulting in difficulties in detecting anomalies amidst benign packets and connections. In addition, the heterogeneity of devices in IoT systems often utilize various protocols and exhibit unique behaviors, making it challenging to have a one-size-fits-all security standard. Furthermore, constrained hardware, such as limited memory, processor, and storage, along with limited power resources, make many IoT devices incapable of having robust detection and recovery functionalities.

\subsection{Explanation of MitM attacks and their types}
\label{sec:Explanation of MitM attacks and their types}

Man-in-the-Middle is a significant threat from a cybersecurity perspective. Before delving into the details of MitM attacks, it's important to understand the aspects affected by this threat. A commonly used model to measure the facets of a threat is the Confidentiality, Integrity, Availability (CIA) Triad \cite{b35}. According to this triad, the impacts of threats are categorized into three separate domains. As previously mentioned, DDoS attacks directly affect the availability of a system. However, MitM attacks directly affect all three sides of the triad, namely availability, integrity, and confidentiality. The triad also implicitly involves privacy, which encompasses not only confidentiality but also authentication and authorization. Therefore, as depicted in Figure 5, MitM is considered a direct privacy-security threat.

\begin{figure}[htbp]
\centerline{\includegraphics[width=0.49\textwidth]{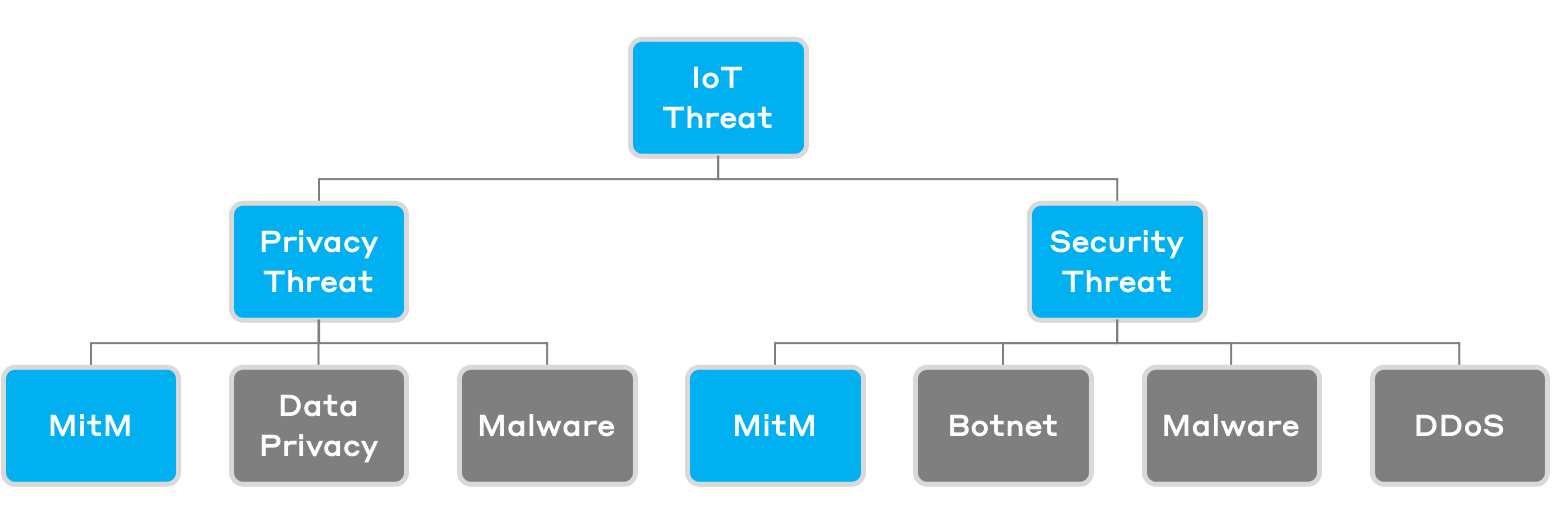}}
\caption{Privacy and Security Threats in IoT}
\label{fig}
\end{figure}

In terms of assets, it is possible to classify security issues into users, data, and infrastructure to examine the effects of attacks on each of them in a system. As for MitM, this attack has a direct influence on users, ranging from their privacy to their physical and mental health. Furthermore, this attack does not only impact data in motion directly, but it impacts data in rest and data in use. In addition, MitM can disturb and ruin network infrastructure, especially in IoT.

\begin{figure}[htbp]
\centerline{\includegraphics[width=0.49\textwidth]{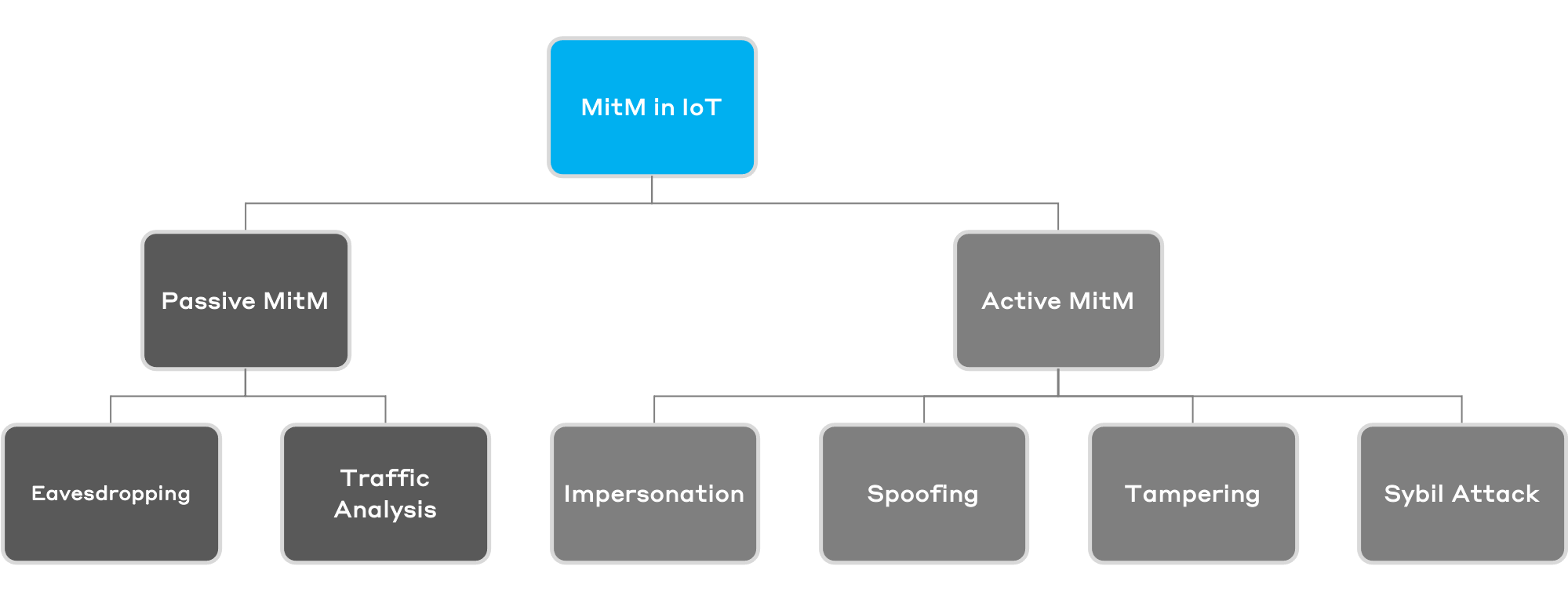}}
\caption{Passive and Active Man-in-the-Middle Attacks}
\label{fig}
\end{figure}

Overall, as shown in Figure 6, it is possible to divide MitM attacks into two distinct categories, which are passive and active attacks: 

\begin{itemize}
    \item A passive man-in-the-middle (MitM) attack occurs when an attacker does not change the communication actively. Instead, the attacker covertly listens to the communication to gain access to sensitive information. Even though the attacker is not manipulating the communication, they can still obtain sensitive information like usernames, passwords, and other confidential data, as most traffic is not encrypted properly, according to the introduction section. Eavesdropping is a type of passive MitM.
    \item An active man-in-the-middle (MitM) attack occurs when an attacker intentionally intercepts and modifies the communication between two entities. Once the attacker is in the communication channel, they can manipulate the communication by intercepting, changing, or inserting new messages. Impersonation, spoofing attacks \cite{b30}, and data tampering are the most common types of active MitM.
\end{itemize}

One of the essential points that should be mentioned here is that MitM is a multilayer threat \cite{b16}, which means it overshadows all architecture layers of the IoT. A spoofing attack can prove this assertion. DNS Spoofing affects the application layer, IP Spoofing influences the network layer, and ARP Spoofing acts on the link layer. In addition, perpetrators can implement Rogue Access Point attacks to intercept a network on the physical layer.

\begin{figure}[htbp]
\centerline{\includegraphics[width=0.49\textwidth]{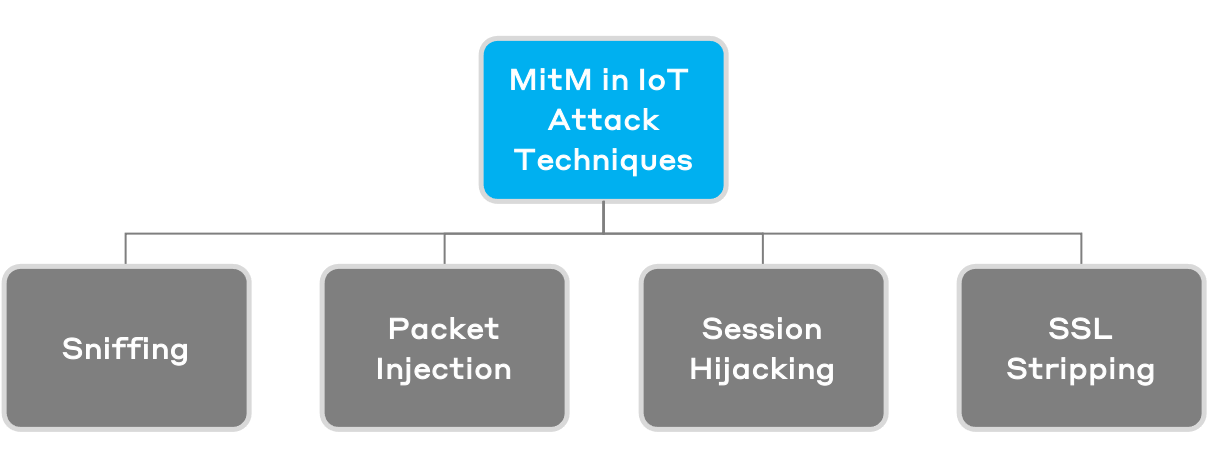}}
\caption{Some Techniques of Man-in-the-Middle}
\label{fig}
\end{figure}

Alongside the types of Man-in-the-Middle attacks described above, several techniques can be used to implement these attacks. As shown in Figure 7, this part briefly explains the common techniques of MitM \cite{b40} to shed light on the subject.
\begin{itemize}
    \item Sniffing refers to the interception and analysis of network traffic to obtain sensitive information between two devices. When it comes to IoT, attackers may utilize this method to gain access to unencrypted data transmitted among IoT devices and also between IoT devices and cloud services. Wireshark \cite{b41} and Cain and Abel \cite{b42} are two valuable tools to execute this technique.
    \item Packet injection is a technique that permits attackers to manipulate network traffic by adding harmful packets to the network. In the realm of IoT, attackers may utilize this technique to interfere with the communication between IoT devices and cloud services or to inject commands that can modify the behavior of the devices. Scapy \cite{b43} and MITMf \cite{b44} are powerful tools for deploying this technique.
    \item Session hijacking is a method in which an attacker steals a user's session information, enabling them to impersonate the user and access their data. In the realm of IoT, an attacker may use session hijacking to gain access to the user's IoT devices and manipulate them remotely. Bettercap \cite{b45} is one of the most used, powerful tools to implement this technique.
    \item SSL stripping is a technique that involves converting an encrypted SSL connection to an unencrypted one. When it comes to IoT, attackers can utilize SSL stripping to intercept and alter data that is exchanged between IoT devices and cloud services in a stealthy manner. SSLStrip \cite{b46} is one of the well-known tools to perform this technique. 
\end{itemize}

\subsection{How MitM target IoT devices and their impact}

There are generally three major steps to execute a Man-in-the-Middle attack in an IoT environment, ranging from simple eavesdropping attacks to complex tampering attacks. The first step involves the adversary entity conducting reconnaissance of their target setting. The second step involves taking advantage of potential vulnerabilities to intercept targets, and the final step involves modifying trading data. These steps can be carried out using respective tools.

\begin{enumerate}
    \item Network scanning and targeting: To execute a MitM attack on an IoT device, the attacker needs to first discover the IP address of the target device by scanning the network. This can be accomplished using tools like Nmap \cite{b47} or search engines such as Shodan \cite{b48}, which can locate internet-connected devices.
    \item Interception and traffic analysis: Once the target device is found, the hacker can utilize network traffic analyzers like MITMf, and Bettercap tools to intercept the communication taking place between the devices. Following that, they can inspect the traffic using packet-capturing tools like Wireshark and Tcpdump \cite{b49} to identify weaknesses in the communication protocol, such as unencrypted data, weak encryption, or other vulnerabilities that may be susceptible to exploitation. Moreover, in IoT settings, lightweight brokers such as Mosquitto Broker \cite{b50} are available as open-source MQTT brokers. These brokers can be utilized for MitM attacks, allowing the attacker to intercept and modify MQTT messages before sending them to their desired destination.
    \item Data modification and vulnerability exploitation: Upon identifying any vulnerabilities, the hacker can alter the transmitted data using tools such as Mallory \cite{b51}, which could involve injecting malevolent code or stealing confidential data. In the event that the hacker has effectively tampered with the data, they can exploit the vulnerabilities to gain illicit access to the device or to steal data with the help of tools like Metasploit \cite{b52}. This step describes active MitM attacks.
\end{enumerate}

\subsection{Examples of real-world MitM in IoT networks}

Reviewing recent real-world Man-in-the-Middle attacks can enable security specialists to gain a comprehensive understanding of the drawbacks and potential methods to resolve the issues. Therefore, the following are some recent and relatively well-known MitM attacks that will be briefly discussed:

In late 2022, consumer brand Eufy came under public fire as they failed to patch massive security vulnerabilities to their cameras and lied about local-only video storage claims. It was found that hackers could intercept the video feed and access systems with privileges using buffer overflow or a MitM attack \cite{b18, b19}.

In March 2021, a MitM attack was performed on Verkada's security camera systems. The attackers, being part of the hacker collective APT-69420, managed to put their hands on credentials stored on a Jenkins server that allowed them to infiltrate the customers' cameras \cite{b20}.
 
In another instance, the Equifax data breach incident in 2017 is a perfect example of a MitM attack. The credit company neglected to renew a public key they were using in their security systems, which ended up being detected by hackers who took advantage and created multiple web shells, ultimately luring a lot of Equifax customers into fake websites \cite{b21}.

\section{Methodology}

In this section, the methodology that was utilized to locate articles for this paper regarding man-in-the-middle attacks in IoT networks, along with its challenges and solutions, will be presented. The research question posed was, "What are the challenges and potential solutions for preventing and mitigating man-in-the-middle attacks in IoT networks?"

A combination of search terms and databases was utilized to retrieve articles relevant to the research question. This involved identifying key concepts and creating a list of search terms and synonyms for each concept, including "man-in-the-middle attacks," "IoT security and privacy," "IoT networks," "prevention," "security solutions," and others. Conceptually, this paper is divided into three sections: statistical data \hyperref[sec:Introduction]{(mostly in the first section: Introduction)}, definitions \hyperref[sec:Background and Theory]{(mostly in the second section: Background and Theory)}, and challenges and solutions \hyperref[sec:A Comparative analysis of the current state of IoT and MitM attacks]{(especially in the fourth and fifth sections)}. Reputable institutes, companies, and scientific articles were used as references. To locate scientific articles, we selected databases such as IEEE Xplore, ScienceDirect, Springer, Wiley Hindawi, and MDPI. These databases were chosen based on their reputation for providing high-quality research articles and extensive coverage of the topic. The selection of articles for this paper involved the application of specific criteria. Mostly peer-reviewed journal articles and conference papers were considered, ensuring that experts had evaluated them. To guarantee up-to-date and relevant information, our search was limited to articles published within the past ten years. The inclusion criteria for articles comprised English language and online accessibility.

After compiling a list of possible articles, we assessed their suitability for our paper by examining various factors such as the author's expertise, publication date, and source. We also scrutinized each article's abstract and introduction to determine its relevance to our research question. Challenges and limitations were encountered during the article search, including a limited number of articles on the topic and non-standardized terminology for describing man-in-the-middle attacks in IoT networks.

The methodology used to find articles on man-in-the-middle attacks in IoT networks involved identifying key concepts and search terms, selecting relevant databases, evaluating articles, and acknowledging challenges and limitations. Despite these challenges, several high-quality articles were located that provided valuable insights into preventing and mitigating man-in-the-middle attacks in IoT networks.

\section{A comparative analysis of the current state of IoT and MitM attacks}
\label{sec:A Comparative analysis of the current state of IoT and MitM attacks}

\subsection{Vulnerabilities of IoT devices}
\label{sec:Vulnerabilities of IoT devices}

IoT devices, being composed of mainly three layers of architecture as mentioned by \cite{b17}, have different types of vulnerabilities at each layer. The application layer is prone to data access and authentication security issues, the network layer is vulnerable to compatibility issues as well as privacy and cluster security problems, and the link, physical and perception layer is inclined to any node-related attack, such as war driving, node capture, fake node, mass node authentication, etc. Here is a list of some of these attacks, subdivided by each layer:

\begin{enumerate}
    \item Application Layer:
    \begin{enumerate}
        \item DNS spoofing attacks, as mentioned previously in \hyperref[sec:Explanation of MitM attacks and their types]{section II subsection B}, is one of the most common types of active MitM attacks. They consist of intercepting the communication between a user and the server, changing the IP address linked to the domain name requested by the user, and providing a fraudulent website instead. This attack will be explained in detail in the following section.
        \item Session hijacking is a cyber-attack where the attacker takes control of a valid session between an IoT device and its connected application or server, allowing them to manipulate data or actions carried out by the IoT device. \cite{b57} It is considered a variant of a Man in the Middle Attack, but with a slightly bigger severity because the attacker takes over the application or the system instead of simply stealing or modifying some internet packets.
    \end{enumerate}
    \item Transport and Network Layer:
    \begin{enumerate}
        \item IP spoofing is an attack in which the source IP address of an IoT device is falsified, creating the illusion of a legitimate communication. It enables interception, manipulation, or injection of malicious data into the network. It can also lead to the impersonation of a trusted entity in the communication between the IoT devices. This attack will be explained in the following section.
        \item Hello flooding is a type of DoS attack where the attacker exhausts node resources by sending multiple requests subsequently or at the same time.\cite{b16} This specific attack takes advantage of the routing protocols, which require nodes transmitting the HELLO packets to communicate with their neighboring nodes in the network. Therefore, when an intruder impersonates a node in the system, it can easily build communications with the rest of the nodes by convincing them that it's their neighbor and proceeding with the flooding.
        \item SSL stripping is a type of attack in which a perpetrator intercepts the communication between IoT devices that employ SSL or TLS encryption and forcibly downgrades the connection to an unencrypted version.\cite{b57}
        \item Routing table poisoning is a form of attack in which an adversary manipulates network layer tables to redirect communication between IoT devices to a malicious destination, enabling interception, modification, or blocking of communication. It can result in unauthorized access, service disruption, or data leakage/disclosure.
    \end{enumerate}
    \item Link, Physical and Perception Layer:
    \begin{enumerate}
        \item ARP poisoning is a cyber-attack in which an attacker manipulates the ARP tables at the data link layer of an IoT system.\cite{b30} This manipulation involves associating the attackers' MAC address with the IP address of legitimate devices to deceive and obtain unauthorized access. This attack will be explained in the following section.
        \item An exhaustion attack is a type of DoS attack where by constantly attacking the network, the batteries of the IoT devices get exhausted and deactivated. As mentioned in \cite{b33}, a collision between machines' communications through MAC protocols results in repeated attempts at re-transmission, which highly drains the battery resources. This will cause the devices' batteries to die, leading to exhaustion in the end nodes.
        \item Hardware implants, sometimes called hardware Trojans \cite{b59}, are physical devices that attackers can secretly plant in IoT devices. Such implants enable the IoT devices to intercept, manipulate or inject malicious content into their communications. Implants can be concealed in different parts of a device, such as cables, connectors, or even circuit boards.
    \end{enumerate}
\end{enumerate}

\subsection{Analysis of MitM attacks on IoT devices}
\label{sec:Analysis of MitM attacks on IoT devices}

As the Man in the Middle Attack is a multi-layer threat \cite{b16}, and spoofing is considered the most common of the MitM attacks \cite{b31}, in this subsection, we will go over the various spoofing attacks that exist at each layer:

\begin{enumerate}
\item
Domain Name System (DNS) Spoofing is a MitM attack at the application layer. The DNS server is responsible for translating the commonly known domain name to the numerical IP address. This IP address will create a communication route between nodes to transfer the requested information \cite{b61}. If the domain name doesn't create a direct link to the IP address, it will call another server that might have that information, until it is able to create a list of connected server names that will be able to connect the initial domain name to the numerical IP address. This list of translations will be cached on the users' computer in case the same address is evoked in the near future. During a DNS MitM attack (as shown in Figure 8), a hacker alternated the IP address related to a domain name, thus providing false information to the user who then proceeded to cache the fake DNS entry.

\begin{figure}[h]
\centering
\includegraphics[width=0.49\textwidth]{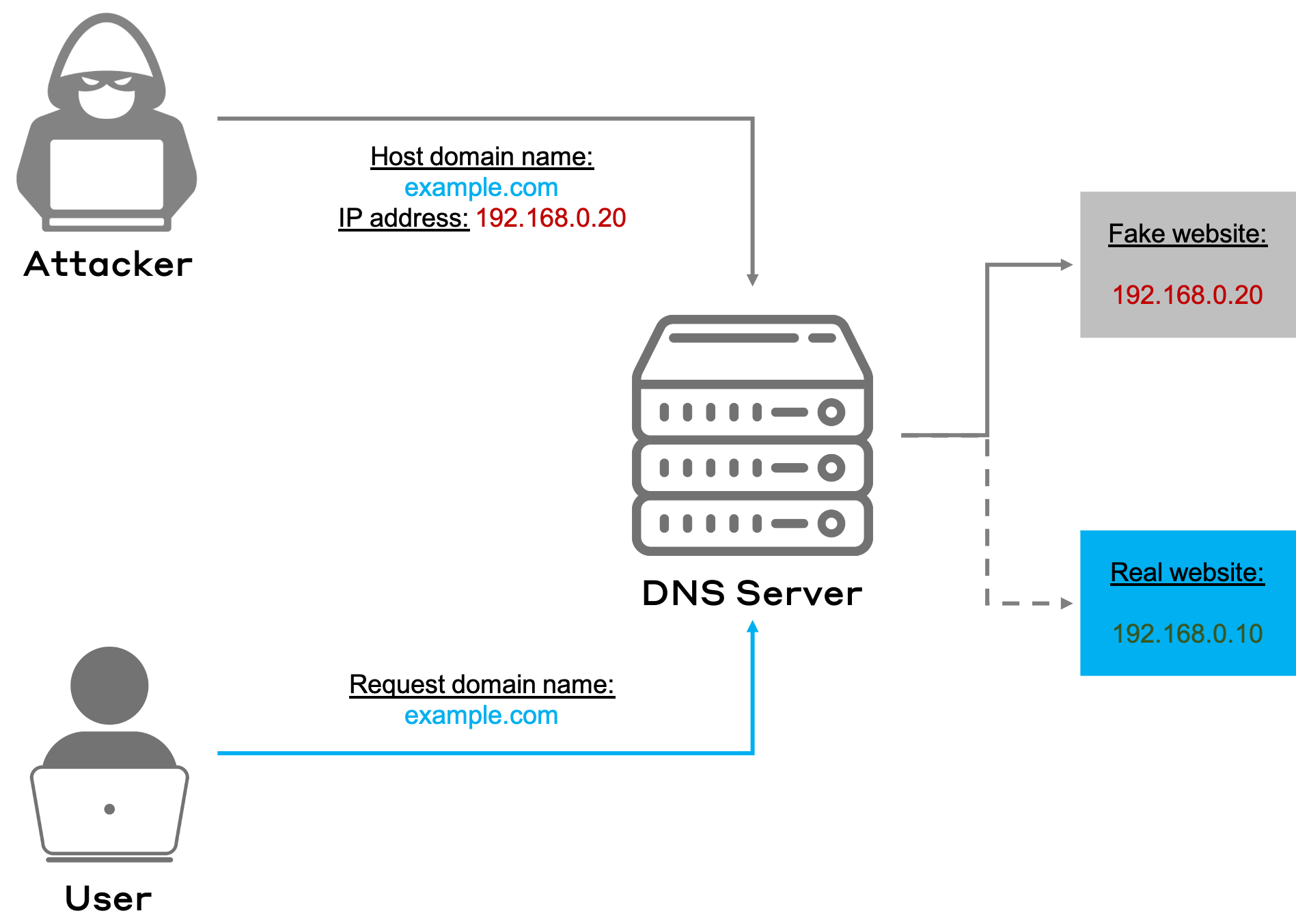}
\caption{Example of DNS Spoofing}
\label{greedyRdmVsSorted}
\end{figure}

\item 
IP Spoofing is a MitM attack at the network layer. During an IP spoofing attack (as shown in Figure 9), the hacker intercepts the transmission of a packet and modifies the IP source address in the header of the packet to convince the user that the packet arrives from a legitimate source \cite{b62}. Once the connection is created between the hacker and the victim's IP address, the communication channels are open for the hacker to induce its victim to go on harmful and fake websites.

\begin{figure}[h]
\centering
\includegraphics[width=0.49\textwidth]{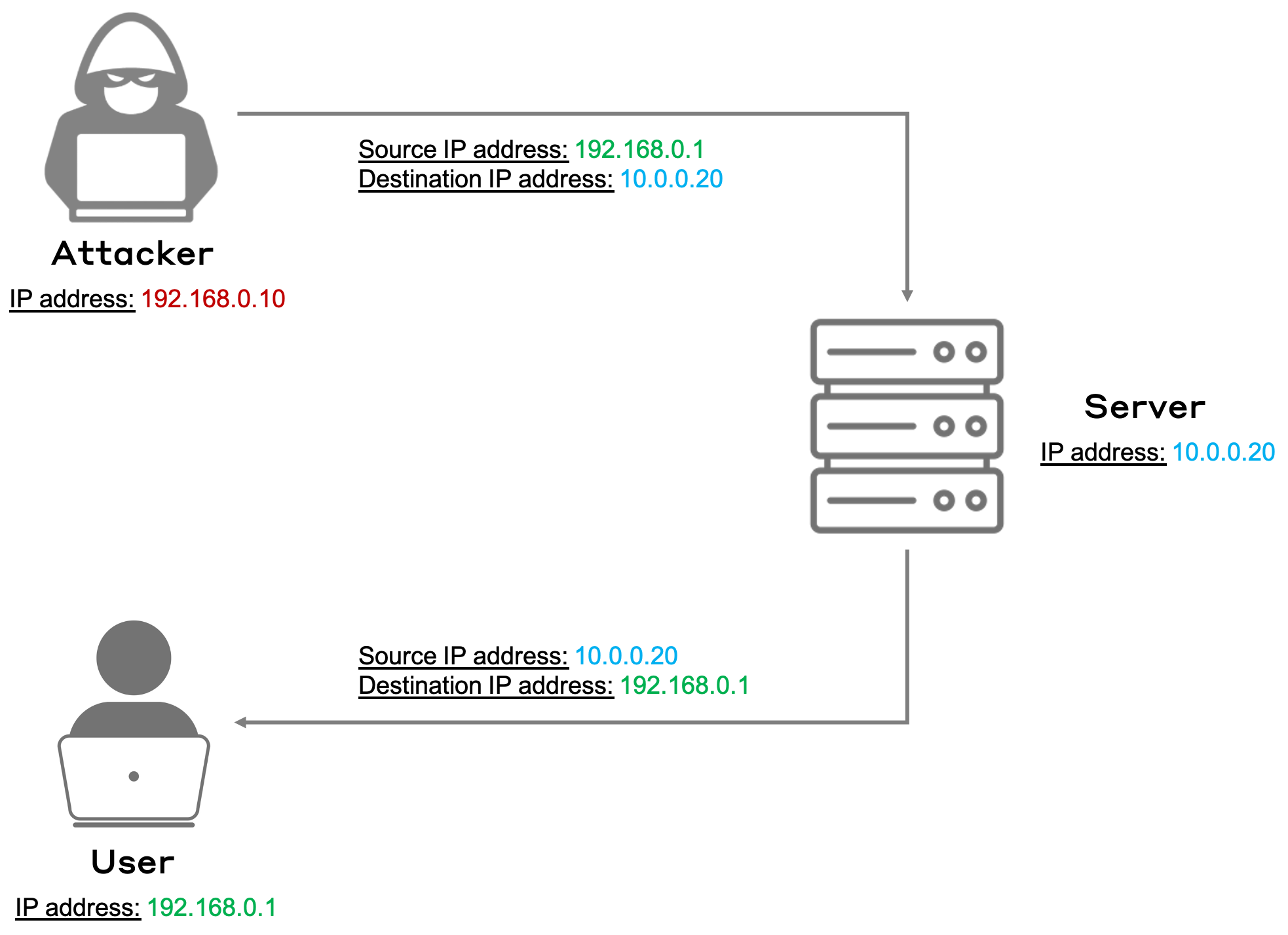}
\caption{Example of IP Spoofing}
\label{greedyRdmVsSorted}
\end{figure}

\item 
Address Resolution Protocol (ARP) Spoofing is a common MitM attack at the perception (or link) layer. ARP is a stateless protocol that is used by a local machine to map the MAC address of the local area network (LAN) to the IP address of that network. This is a weak protocol because it has no authentication method, making it a prone environment for a MitM attack. Such an attack is executed when the hacker sends an ARP reply with the legitimate IP address of the host the machine is trying to reach (as shown in Figure 10), but its own MAC address instead of the actual address of the host. Therefore, the victim will receive the ARP reply and will contain the wrong information linking the right IP address to the wrong MAC address \cite{b63}. For example, in the case where the user of the local machine tries to reach a host database, if a MitM attack occurs at the ARP level, the user will end up sending its IP packets to the attacker instead of the database. This is called ARP poisoning. \cite{b30}

\begin{figure}[h]
\centering
\includegraphics[width=0.49\textwidth]{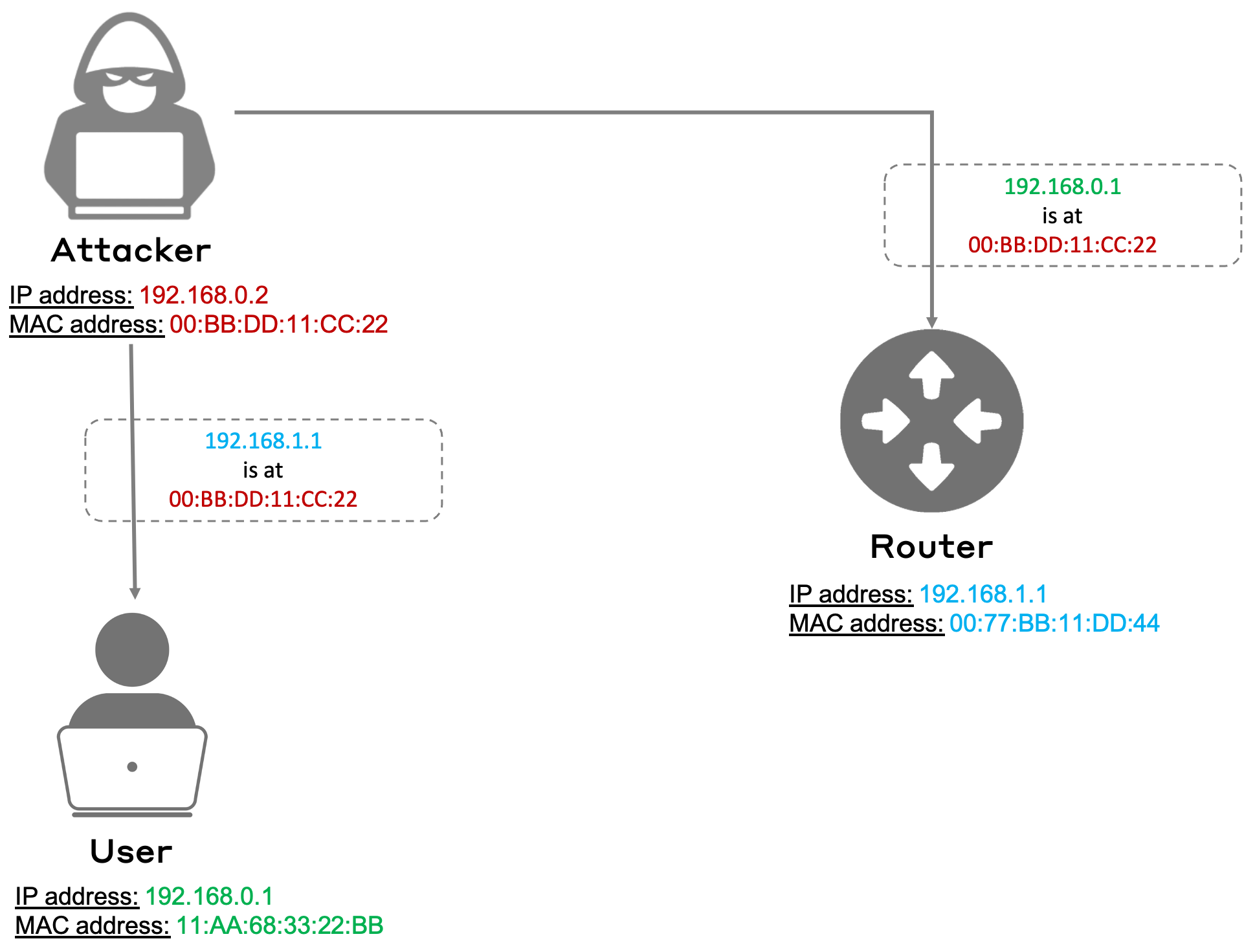}
\caption{Example of ARP Spoofing}
\label{greedyRdmVsSorted}
\end{figure}
\end{enumerate}

\subsection{Current prevention techniques}
\label{sec:Current prevention techniques}

From protocols to machine learning, there are multiple techniques being used to prevent attacks on IoT devices. 
First, there are multiple standardized protocols already in place that ensure basic security measures when implementing IoT devices. Based on this survey \cite{b16}, here are the main protocols per each layer:
\begin{enumerate}
\item Constrained Application Protocol (CoAP) is an ultralight application level protocol. It is mainly used in Machine-to-Machine (M2M) applications on a low-power and low-bandwidth constrained network.
\item The Datagram Transport Layer Security (DTLS) protocol is used at the transport layer for situations requiring fast data transfer with short response times (such as video or game rendering). This protocol secures communication between clients and servers by using certification-based authentication methods \cite{b64}.
\item IPv6 is the main Internet key protocol for IoT devices and has been considered the main Internet Standard since 2017 \cite{b32}. This protocol is behind any communication at the network layer between IoT devices.
\item IEEE 802.15.4 is the protocol used in Bluetooth, WiFi, or WLAN, which are common physical networks. 
\end{enumerate}

Nevertheless, these protocols are simply standard guides directing the way communication should function through networks. Such protocols should be followed to avoid being intercepted by intruders, yet it doesn't prevent all types of attacks from happening. Therefore, here are different specialized prevention techniques that are used for the various attacks listed in \hyperref[sec:Vulnerabilities of IoT devices]{section IV subsection A}:

\begin{enumerate}
    \item Application Layer:
    \begin{enumerate}
        \item Using Domain Name System Security Extension (DNSSEC) enhances DNS security with digital signatures, verifying data integrity and authenticity to prevent DNS spoofing attacks and malicious data injection into DNS caches. \cite{b53}
        \item Ensuring secure session management with unique session IDs and regular password rotation can prevent session hijacking attacks.
    \end{enumerate}
    \item Transport and Network Layer:
    \begin{enumerate}
        \item Using TLS and HTTP Strict Transport Security (HSTS) \cite{b60}, which is a security feature that allows a website to specify that it should only be accessed over HTTPS and not over HTTP.
        \item Implementing Role-Based Access Control (RBAC) to restrict user permissions within a network based on roles (e.g. admin vs regular user), utilizing encryption to safeguard stored data, and employing data integrity checks (e.g., checksums, hash functions, digital signatures) to detect any data alterations.
        \item Using specialized tools and techniques like Reverse Path Forwarding (RPF) and Unicast Reverse Path Forwarding (uRPF)\cite{b54}, as well as Network Intrusion Detection and Prevention Systems (NIDS/NIPS), to thwart IP spoofing attacks.
        \item Using deep learning (DL) and optimization algorithms can help mitigate Hello flooding attacks. \cite{b58}
    \end{enumerate}
    \item Link, Physical and Perception Layer:
    \begin{enumerate}
        \item An exhaustion attack can be avoided by using timers or a rate limitation on the number of requests sent to an IoT device. \cite{b16}
        \item Enabling port security can limit MAC addresses, preventing unauthorized devices from connecting and protecting against ARP spoofing attacks. Also, deploying network monitoring and intrusion detection systems (IDS) to detect ARP spoofing by reviewing logs and analyzing network traffic for signs of ARP spoofing attempts.
        \item Network monitoring detects suspicious network traffic that may indicate hardware implants, such as unusual data flows, anomalies, or unauthorized access attempts. Hardware integrity verification systems let us check components for unauthorized modifications using techniques like hardware integrity checking, trusted platform modules (TPM), or hardware security modules (HSM) \cite{b55}.
        \item Monitoring for suspicious activity using wireless intrusion detection and prevention systems (WIDS/WIPS)\cite{b56} by reviewing logs and analyzing network traffic to detect and alert unauthorized devices or attempts to intercept wireless signals. Using string encryption to ensure that all wireless communications are encrypted using robust encryption algorithms. Also, utilizing MAC (Media Access Control) address filtering allows one to specify which devices can connect to a wireless network based on their MAC addresses.
    \end{enumerate}
\end{enumerate}

Finally, Machine Learning (ML) and Deep Learning (DL) are being leveraged to counter some security threats as well, by building models that can predict which behaviors are threatening. \cite{b24} In classifications made by an ML algorithm, there is a chance of false positives and false negatives, making this technique not perfect. Nevertheless, deep learning can be used to determine the accuracy of every prediction, which makes ML and DL together great tools to use for detecting malicious attacks.

\subsection{Current techniques to mitigate and manage MitM attacks on IoT devices}
\label{sec:Current techniques to mitigate and manage MitM attacks on IoT devices}

Multiple prevention techniques exist to avoid MitM attacks on IoT devices: multi-factor authentication, setting up VPNs, using secure communication tools, making sure SSL certificates are secure and systems are updated, etc. On top of that, ML can be leveraged to detect anomalies through unsupervised and supervised learning, as seen in the paper \cite{b28}. Machine Learning and Distributed Ledger Technologies (DLTs) can help improve IoT infrastructure and prevent cybersecurity attacks \cite{b29}. Here are some techniques used in the MitM attacks described previously in \hyperref[sec:Analysis of MitM attacks on IoT devices]{section IV subsection B}:
\begin{enumerate}
    \item DNS Spoofing at the application layer: Deep Packet Inspection (DPI) and Deep Flow Inspection (DFI) are modules that analyze the traffic of a network and determine if it has anomalies. \cite{b34} A DPI performs an analysis of packets components, such as the header and the payload.
    DFI analyzes packets features such as the total bytes flow, the packet count flow, the duration of flow, and the average packet bytes flow and flags whenever there is an abnormal activity. DFI can also support encrypted data, but DPI can't.
    \item IP Spoofing at the network layer: It is hard to determine when an attack has happened at the network level, but network monitoring tools can help determine when the information in the response has been modified. Ingress and egress filtering is one of the methods which can help prevent IP spoofing. 
    \item ARP Spoofing at the link layer: There are multiple existing techniques that prevent ARP spoofing, ranging from static ARP tables that map the right MAC addresses to their corresponding IP addresses to switch security checks which consist of checking each ARP message and filter out messages that seem to be malicious \cite{b24}. There are also Machine Learning algorithms being developed to mitigate this attack. In the article \cite{b23}, the researchers use different classification algorithms to detect an ARP spoofing MitM attack. They used two different IoT devices, a voice recognition device SKT NUGU (NU 100) and a wireless EZVIZ WiFi camera, connected to a wireless network to which other devices were connected as well. A variety of network packet files (pcap) with multiple packets in each, were transmitted on this network, where 6 out of the 42 raw pcap files contained MitM attacks. Each pcap file contained 7 features: packet sequence number, its transmission time, source IP address, destination address, the protocol used, length in bytes and info containing additional details. These packets features were used to train the classification algorithms (Logistic Regression, Random Forest, and Decision Tree), which were then used on the test set, to determine which packets were compromised. The MitM dataset used in this experiment contained 194184 observations, where 52.74\% were attack packets, and 47.53\% were normal ones. The results for each classification algorithm were extremely good, giving 99\% precision, recall, and f1-score for the logistic regression, and a 100\% score for the same metrics for the random forest and decision tree algorithms. Surely, such high scores raise questions such as: what is the quality of the data and how can we assess that it's not biased? Nevertheless, this experiment demonstrates how ML algorithms can be leveraged to detect MitM ARP spoofing attacks. 
    
\end{enumerate}

\section{Open issues}
\label{sec:Open issues}

\subsection{Challenges of prevention}

The growth in the number of connected devices discussed previously has led to the democratization of IoT devices. More and more devices are available to end-users with technological skills ranging from expert-level to tech-beginners. For the latter, this means that the end-user who may not have the technical knowledge or understanding of cybersecurity risks, can become a potential risk. The end-user may inadvertently introduce vulnerabilities into the network by failing to properly secure their devices or failing to keep their software up to date.

For organizations, the deployment of a large number of IoT devices can make them vulnerable to MitM attacks, where attackers intercept and manipulate data sent between devices. Active monitoring becomes a significant challenge, as plaintext communication, open ports, and weak credentials used on these devices can all be exploited by attackers to gain unauthorized access to the network and its large data volume. This is particularly problematic when patching security cameras and printers, where the sheer number of devices creates a window of vulnerability if a security patch is not applied promptly. Not only abundant, these devices are also the most vulnerable ones in an enterprise network \cite{b5}. To prevent MitM attacks, organizations must implement security policies that outline how IoT devices are deployed, managed, and monitored, and ensure that regular assessments are conducted to identify any vulnerabilities that may exist. 

While active monitoring is getting better and better with the use of machine learning in IT security (a subject which will be discussed in the next section), the recognition of malicious patterns is mainly based on previous experience or available datasets. However, generating such training data is quite challenging as the IoT devices environment is itself constantly changing \cite{b26}. Also, finding quality publicly available datasets can be quite the task \cite{b16}. For all these reasons, the prevention of attacks on these devices represents a major challenge that requires a lot of effort from organizations, manufacturers, and end-users, to try to mitigate security risks.

\subsection{Emerging technologies for IoT security}

The IoT security industry currently uses several commercial technologies to secure networks, including asset management and Computerized Maintenance Management Systems (CMMS), Security Information and Event Management (SIEM), Security Orchestration, Automation and Response (SOAR), Next-Generation Firewalls (NGFW), Network Access Control (NAC), and wireless/network management solutions \cite{b5}. While these technologies are effective, they may not always detect sophisticated attacks. Emerging technologies in IoT security, such as intrusion detection using deep learning, show promise in addressing this limitation. Deep Learning algorithms can analyze large volumes of data and identify anomalous behavior patterns, which may indicate a potential security breach. Traditional machine learning methods like Logistic Regression, Random Forest, or Decision Trees already show nice results, as some researchers can get more than 98\% accuracy on a publicly available dataset \cite{b23}. It is important to note that the said dataset is very simple and that such precision is not to be expected in a real-world scenario, as it is mostly used to compare different approaches. However, Deep Learning is especially useful in detecting and responding to zero-day attacks, which exploit previously unknown vulnerabilities in IoT devices \cite{b22}. Current Deep Learning methods used and developed are using Convolutional Neural Network (CNN), Gated Recurrent Unit (GRU), and Long Short Term Memory Neural Network (LSTM) for intrusion detection, other classifiers are being discussed in the research community, such as Genetic Algorithm and Bidirectional Short Term Memory (BiLSTM), with the hope that they will increase performance \cite{b22}. By leveraging emerging technologies such as deep learning, organizations can strengthen their IoT security posture and stay ahead of potential threats.

\subsection{Future trends of MitM attacks and how to start preparing for them}

One of the prevention challenges previously discussed was linked to the growth in the number of connected devices. This has created a challenge due to the absence of a standard architecture for IoT. With billions of objects getting attached to the traditional internet daily, it is essential to have an architecture that is adequate for easy connectivity, communication, and control. To ensure security in IoT, devices should contain an Identity Manager \cite{b15}. Without it, the heterogeneity of the environment gives MitM attackers a perfect ground to find vulnerabilities. It is also much more difficult for the security community as they would have to patch these on a device basis.

MitM attacks on IoT devices that use Long Range Wide Area Networks (LoRaWAN) have significantly increased. Hackers can target these networks using a method called Address Resolution Protocol (ARP) spoofing, which sends fake ARP messages over a LoRaWAN. ARP is a protocol used to associate the network layer address (IPv4) with a physical address, such as a Media Access Control (MAC) address. During an ARP spoofing attack, an attacker sends forged ARP messages to the target's device or router, tricking it into associating the attacker's MAC address with the IP address of the intended destination. As a result, any traffic intended for the destination is sent to the attacker's device instead. This could allow the attacker to eavesdrop on the communication between the two devices without being detected and can be used to steal login credentials, redirect users to a fake website, or launch other types of attacks. One way to prevent these is to use Static ARP Tables. By having static addresses stored, it ensures that any modification would need a manual update of the tables by the user on all the hosts \cite{b24}. It is important to note that this method isn't applicable to particularly large networks with many devices as the manual update would require a great effort.

When faced with a MitM attack on a network, it can be measurably observed that there is an unusual delay, caused by the attacker’s information rerouting. This characteristic can be used as a prevention and detection method. A technique called Hybrid Routing for Man-in-the-Middle (MitM) Attack Detection in IoT Networks was developed to appoint dedicated nodes to route IoT traffic. By selecting devices with enough computational capabilities as nodes, traffic can be routed through them to improve the stability of travel times. Coupled with an inference Algorithm, this would make anomaly detection much easier as a MitM attack would result in an inconsistent transmission time \cite{b27}. 
By implementing these network-wide techniques, the challenge brought on by the heterogeneity of the IoT device landscape could be tackled and could prepare users against potential attacks.

\subsection{The potential impact of new technologies in IoT and MitM attacks}

New technologies in IoT have the potential to significantly impact the security landscape. As more of them are becoming interconnected, the risk of MitM attacks is increasing. To mitigate this risk, it is essential that new IoT technologies include robust security features that can not only detect, but also prevent MitM attacks.

It is also important to keep in mind the impact of compatibility and scalability as the lack of standardization in the IoT industry can create compatibility issues and security risks. The development of new standards that ensure interoperability, scalability, monitoring, and attack prevention is essential to ensure the widespread adoption of new IoT technologies \cite{b25}.

Regulation is also a critical consideration for the IoT industry, as IoT devices collect and transmit sensitive data. There is a need for regulations to ensure the security of IoT devices and to protect users from potential harm caused by breaches and leaks. The main issue is that the laws and public guidelines are created and updated at a much slower pace than the technologies they are governing. Reacting slowly to innovation could have negative impacts on every party whether it is on a consumer, an industrial member or even a whole state.

Overall, the development of new IoT technologies must consider the potential impact on security and privacy, compatibility, and regulation to ensure the safe and widespread adoption of IoT. By prioritizing these considerations, IoT can continue to grow and expand, offering increased connectivity, convenience, and efficiency in various applications.

\section{Conclusion}

IoT devices are gaining popularity amongst technology professionals and regular individual end-users. Currently, the security of such devices is crucial, but lacking. They are subject to many MITM attack vectors due to the heterogeneity of the IoT space. These attacks are considered direct privacy and security threat. Many current methods are used to prevent and combat such attacks but have some limitations. Some of these limitations are the complexity brought on by the increase in network size, the limited compute power and battery capacity of IoT devices, and the simplicity of the data sets used to train Machine Learning and Deep Learning models, as they do not accurately mimic real-world use cases. In the short-term future, techniques using Deep Learning will be explored and used to detect intrusions. As for the industry in general, it is important to prioritize security and privacy, but also compatibility, regulation, and the integration of emerging technologies to improve active monitoring and detection of potential security breaches. In terms of future research, Network Architecture Optimization or Standardization for Hybrid Routing can be interesting to investigate to try to adapt the technique to larger network sizes. Other broader avenues would be to help other researchers with data set creation to mimic larger size networks, and finally, the use of blockchain/DLTs as an immutable decentralized database can be explored.


\end{document}